\documentclass{elsart5p}

\usepackage{graphicx}
\usepackage{amssymb}

\newcommand{\ybco}{YBa$_2$Cu$_3$O$_{6+\delta}$}
\newcommand{\ybcoa}{YBa$_2$Cu$_3$O$_{6.45}$}

\newcommand{\lsco}{La$_{2-x}$Sr$_x$CuO$_4$}
\newcommand{\bscco}{Bi$_2$Sr$_2$CaCu$_2$O$_{8+\delta}$}
\newcommand{\lbco}{La$_{2-x}$Ba$_x$CuO$_4$}
\newcommand{\lbcoo}{La$_{15/8}$Ba$_{1/8}$CuO$_4$}
\newcommand{\ccoc}{Ca$_{2-x}$Na$_x$CuO$_2$Cl$_2$}

\begin{document}
\begin{frontmatter}


\journal{SCES '08}


\title{
Photoemission signatures of valence-bond stripes in cuprates:\\
Long-range vs. short-range order
}

%
%
%
%
%
%
\author[K]{Alexander Wollny}
\author[K]{and Matthias Vojta\corauthref{1}}

%

\address[K]{
Institut f\"ur Theoretische Physik,
Universit\"at zu K\"oln, Z\"ulpicher Stra{\ss}e 77,
50937 K\"oln, Germany.
}

%
%
%
%


%
%
%
%

\corauth[1]{Email: vojta@thp.uni-koeln.de}


\begin{abstract}
Recent experiments indicate that the tendency toward the formation
of unidirectional charge density waves (``stripes'') is common
to various underdoped cuprates.
We discuss momentum-resolved spectral properties of valence-bond stripes,
comparing the situations of ideal and short-range stripe order,
the latter being relevant for weak and/or disorder-pinned stripes.
We find clear signatures of ordered stripes, although matrix element effects
suppress most shadow band features.
With decreasing stripe correlation length, stripe signatures are
quickly washed out, the only remaining effect being a broadening of antinodal
quasiparticles.
This insensitivity of photoemission to short-range stripe order may be
employed to distinguish it from nematic order, e.g. in underdoped
\ybco.
\end{abstract}

%
%



\end{frontmatter}


\section{Introduction}
\label{sec:intro}

States with broken lattice symmetries are of vital interest
in the field of high-temperature superconducting cuprates,
for at least two reasons:
Such states reflect interesting physics inherent to doped Mott insulators,
and they may compete with $d$-wave superconductivity.
One established example is the variety of unidirectional charge-density-wave (CDW) states,
commonly dubbed stripes \cite{pnas,stevek,antonio,jan,jt95},
which break both translation and rotation symmetries of the lattice.
Another candidate is a nematic phase, where only rotation symmetry is broken.
The melting of a stripe phase can proceed in a single transition
or via an intermediate nematic phase \cite{pnas,stevek}.

Experimentally, significant progress has been made over the last few years
in determining the properties of stripe states.
Originally, static magnetic stripe order was detected in neutron scattering
in \lsco\ (214) compounds \cite{jt95,lsco,waki}.
Later, the simultaneous presence of static charge order was proven by
inelastic x-ray scattering \cite{abbamonte,kim_xray,fink}.
Stripe order was found to be particularly stable near 1/8 doping,
with a spatial period of four lattice spacings in the charge sector.
In other cuprate families, scanning tunneling microscopy (STM) experiments \cite{kapi,hana,ali,kohsaka}
detected spatial modulations (on the surface), with a period close to four.
This static order is short-ranged, locally breaks the lattice rotation
symmetry, and has the strongest signal on the bonds (instead of sites) of the
square lattice of Cu atoms \cite{kohsaka}.

On the theory side, charge segregation in doped Mott insulators had been in fact
anticipated \cite{za89}, and a plethora of papers on the problem appeared
after the initial experimental observations (see Refs.~\cite{stevek,antonio,jan} for
reviews).
Recent experimental results allowed to constrain theoretical stripe models.
For instance, the magnetic excitation spectrum of stripe-ordered \lbcoo\ \cite{jt04}
could be nicely described by a simple model of coupled spin ladders,
pointing toward bond-centered stripes dominated by local singlet
formation \cite{MVTU}.
The real-space structure of such valence-bond stripes has been argued \cite{mv08}
to be consistent with the STM data on underdoped \bscco\ and \ccoc\ \cite{kohsaka}.
A plausible scenario for these materials is that stripes would be
slowly fluctuating in the absence of disorder, but get pinned (and hence static)
due to impurity effects \cite{stevek,maestro}.
Importantly, impurities act as random field on charge stripes,
which leads to glassy short-range order.
Even in compounds with well-ordered stripes,
impurities will limit the low-temperature correlation length and
smear out the finite-temperature ordering transition in the charge sector
(analogous to the two-dimensional random-field Ising model),
as indeed observed very recently in Eu-doped \lsco\ \cite{fink}.
An obvious question is whether other compounds like \ybco, where
clear-cut stripe signatures have not been observed, also display
a tendency toward CDW.
Theoretically, it has been shown that slowly fluctuating (or disordered)
charge stripes influence the spin excitations and induce a characteristic
hour-glass shaped spectrum \cite{vvk}.
Such spectra have been measured in \ybco\ \cite{hayden}, indicating the possibility
that underdoped \ybco\ is close to stripe ordering as well.
Interestingly, neutron scattering in detwinned \ybcoa\ has detected signatures of
spontaneous breaking of lattice rotation symmetry \cite{nematic}, which have a
natural interpretation in terms of nematic order.

Angle-resolved photoemission (ARPES), routinely performed on various cuprates,
has so far not identified unambiguous signatures of stripe order,
and early experimental data \cite{zhou99} remained controversial \cite{valla,claesson}.
In this paper, we shall discuss whether ARPES is suited to detect stripe formation and
to distinguish between tendencies toward stripe and nematic states.
While various papers on this subject are in the literature \cite{salkola,seibold,granath,orgad,millis},
the recent advances in comparing theory and experiment put more constraints
on phenomenological stripe models, see Sec.~\ref{sec:input}.
We shall employ a mean-field model of valence-bond stripes, and also include the influence of
spatial disorder using a proper order-parameter field theory.
For perfectly ordered stripes, our results show multiple bands, gaps, and weak shadow bands,
in agreement with earlier work, and we identify the most prominent features.
However, most traces of stripe order are difficult to discern
if both horizontal and vertical stripes contribute to the signal.
Spatial stripe fluctuations further smear out the signal,
such that damping of antinodal quasiparticles (QP) may the only visible effect
of short-range stripes.
Even for unidirectional stripes (i.e. in the presence of a lattice anisotropy),
the breaking of rotation symmetry is ``weak''.
We propose to use this property for an experimental distinction
between symmetry breaking driven by stripes and driven by nematic order,
applicable e.g. to underdoped \ybco.


\section{Experimental input}
\label{sec:input}

Our phenomenological approach is guided by experimental input.
Let us list the most important results and together with the assumptions
entering our modelling.

(i) We shall assume that the order has stripe instead of checkerboard character,
i.e., locally either horizontal or vertical stripes occur -- this is consistent
with both neutron and STM data.
(ii) Over a significant doping range, the charge order has a period of approximately
four lattice spacings \cite{abbamonte,fink,kohsaka},
and we shall restrict the calculations to ordering wavevectors $K_x=(\pm\pi/2,0)$,
$K_y=(0,\pm\pi/2)$.
(iii) The typical modulation amplitude in the charge sector is $\pm20\ldots30\%$,
according to both STM data on underdoped \bscco\ and \ccoc\ \cite{kohsaka}
(if the contrast in the tunneling asymmetry is interpreted as density modulation)
as well as x-ray data on \lbco\ \cite{abbamonte}.
(iv) We assume bond-centered instead of site-centered stripes, based on
the STM data \cite{kohsaka} and on the modelling of neutron scattering data \cite{jt04}
on \lbco\ \cite{MVTU}.
(v) We assume that the dominant modulations are on the Cu-Cu {\em bonds} \cite{kohsaka};
in a one-band model this translates into modulations of bond kinetic, magnetic, and
pairing energies (all being invariant under spin rotation and time reversal).
In such a case, the bond modulations will have a $d$-wave-like form factor
\cite{mv08,MVOR}.
(vi) We assume that dimerization and bond order are the driving forces behind
stripe ordering \cite{vs}, whereas magnetic long-range order is less important.
For simplicity, we shall model stripe states without static magnetism.
Note that this does not mean that we ignore local-moment physics,
but instead we assume that those moments form singlet valence bonds,
which we account for by modulated hopping within our mean-field theory.
Additional spin order will only have a weak influence on our results
(on the mean-field level), as estimated in Sec.~\ref{sec:model} below.

Finally, in-plane anisotropies need to be discussed.
In compounds without anisotropy between the $a$ and $b$ axes of the CuO$_2$ planes,
horizontal and vertical stripes are energetically equivalent.
Then, domain formation is expected, and ARPES will average over both
stripe directions.
Exceptions occur in detwinned \ybco\ (due to the presence of CuO chains)
and in the LTT phase of 214-compounds, i.e., in \lbco and Nd- and Eu-doped
\lsco.
In the latter cases, stable stripe order occurs, with correlation lengths of
order 100 unit cells \cite{fink}.
However, the preferred stripe direction alternates from layer to layer,
and an experiment probing multiple layers will effectively again average
over both stripe directions.
In detwinned \ybco, the symmetry is orthorhombic, which offers the possibility
to study bulk properties of unidirectionally ordered phases.


\section{Mean-field stripe model}
\label{sec:model}

We employ a mean-field model for striped superconductors,
which is combined with an order-parameter field theory for the collective
CDW degrees of freedom. Both model and methodology have been discussed
in Ref.~\cite{mv08}, and will be sketched here only briefly.

Single-particle properties are calculated from a BCS model of fermions on
the square lattice of Cu atoms:
\begin{eqnarray}
\mathcal{S}_c &=& \int \! d\tau \! \sum_{{\vec k}} \left[
\bar{c}_{{\vec k}\sigma} (\partial_\tau\!+\!\epsilon_{\vec k}\!-\!\mu) c_{{\vec k}\sigma} +
\Delta_{\vec k} (c_{{\vec k}\uparrow} c_{{-\vec k}\downarrow} + c.c.)
\right]
\nonumber\\
\label{sc}
\end{eqnarray}
where summation over spin indices $\sigma$ is implied.
The single-particle dispersion consists of hopping to
first ($t$), second ($t'$), and third ($t''$) neighbors, and
$\mu$ is the chemical potential.
The pairing is of $d$-wave type, $\Delta_{\vec k} = \Delta_{0x} \cos k_x + \Delta_{0y} \cos k_y$
with $\Delta_{0x}=-\Delta_{0y}=\Delta_0$.

\begin{figure}
\centering
\includegraphics[width=2.8in]{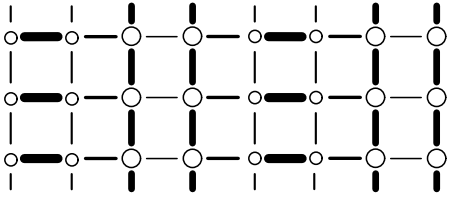}
\caption{
Schematic real-space structure of a valence-bond stripe state with
a $4\times 1$ unit cell.
Cu lattice sites are shown as circles, with their size
representing the on-site hole densities. The line strengths indicate the
amplitude of bond variables like kinetic and magnetic energies.
Our mean-field model, Eq.~(\ref{eq:scpsi}), implements spatial on-site
variations of the chemical potential (via $\kappa_1$) and
spatial variations of nearest-neighbor hopping and pairing amplitudes
(via $\kappa_{2\ldots5}$).
}
\label{fig:schem}
\end{figure}

CDW order is represented by two order-parameter fields $\psi_{x,y}({\vec r}, \tau)$
for horizontal and vertical stripes at wavevectors ${\vec K}_{x,y}$,
such that the real field
$
Q_x ({\vec r} ) = {\rm Re}\,\psi_x ({\vec r}) e^{i {\vec K}_x \cdot {\vec r}}
$
(similarly for $Q_y$) measures the modulation of both the
charge density and bond order (i.e., kinetic energy or pairing amplitude),
for $\vec r$ on sites and bonds, respectively.
The coupling between fermions and the collective CDW fields $Q_{x,y}$ reads
\begin{eqnarray}
\mathcal{S}_{c\psi} &=& \int d\tau \sum_i \Big[
\kappa_1 Q_x({\vec r}_i) \bar{c}_{i\sigma} c_{i\sigma} \nonumber\\
+&&\!\!\!\!\!\!\!\!
\big(
\kappa_2 Q_x({\vec r}_{i+x/2}) \bar{c}_{i\sigma} c_{i+x,\sigma} \,+\,
\kappa_3 Q_x({\vec r}_{i+y/2}) \bar{c}_{i\sigma} c_{i+y,\sigma} \nonumber\\
+&&\!\!\!\!\!\!\!\!
\kappa_4 {\rm sgn}(\Delta_{0x}) Q_x({\vec r}_{i+x/2}) c_{i\uparrow} c_{i+x\downarrow}
\nonumber\\
+&&\!\!\!\!\!\!\!\!
\kappa_5 {\rm sgn}(\Delta_{0y}) Q_x({\vec r}_{i+y/2}) c_{i\uparrow} c_{i+y\downarrow} \!+\! c.c.
\big) +
[x \leftrightarrow y] \Big].
\label{eq:scpsi}
\end{eqnarray}
The coupling constants $\kappa_{1\ldots5}$ decide about the electronic
struture of the CDW state. As stated above, we assume
bond-dominated stripe order with a $d$-wave-like form factor.
Those are induced by $\kappa_{2\ldots5}$,
with the $d$-wave character encoded e.g. in $\kappa_2=-\kappa_3$.

Perfectly ordered static stripes correspond to $\psi_x={\rm const}$,
$\psi_y=0$ or vice versa.
The commensurate period-4 situation translates into an
$8\times 8$ matrix Hamiltonian ($4\times 4$ in the absence of pairing).
Before continuing, let us estimate modulation amplitudes.
To obtain a kinetic-energy modulation of $\pm25\%$, a $15\ldots20\%$ modulation of
the hopping $t$ is required.
To compare with on-site modulations, this number needs to be multiplied by the number
of neighbors, $z=4$. For $t=0.15$ eV, this results in $z\delta t \approx 0.1$ eV.
For comparison, static magnetism causes a spin-dependent on-site (mean-field) potential
of $\pm z J \langle S \rangle/2$, where $J$ is the exchange constant between the spin-1/2 moments.
The ordered moment, $\langle S \rangle$, in the stripe phases of the 214 compounds
has been found by $\mu$SR to be (at maximum) half of the moment of the undoped
antiferromagnet \cite{stevek}, $\langle S \rangle \lesssim 0.15$.
With $J=0.1$ eV, we obtain an upper bound for the triplet-channel modulation of 0.03 eV
(which would be further reduced by stripe disorder and the resulting frustration).
This is significantly smaller than the modulation in the singlet channel,
providing some justification for neglecting static magnetism.


\begin{figure}[t]
\centering
\includegraphics[width=3.1in]{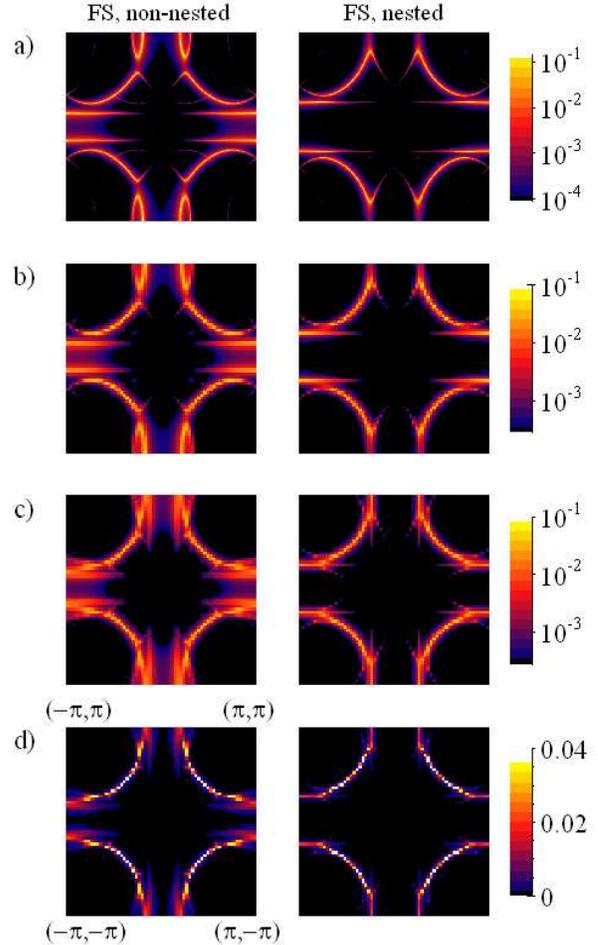}
\caption{
Constant-energy cuts at the Fermi level through the spectral function $A({\vec k},\omega)$
for ideal period-4 valence-bond stripes,
with different chemical potentials leading to Fermi surfaces with
non-nested (left) or nested (right) antinodal pieces.
a) Vertical stripes,
momentum resolution $\pi/256$, energy resolution 2 meV, logarithmic intensity scale.
b) Same as a), but momentum resolution $\pi/32$ and energy resolution 7 meV.
c) Same as b), but horizontal and vertical stripes are added.
d) Same as c), but with a linear intensity scale.
A prominent stripe signature are straight Fermi surface pieces
perpendicular to the stripe direction near the antinodal points.
}
\label{fig:pes1}
\end{figure}

\begin{figure*}[t]
\centering
\includegraphics[width=6.5in]{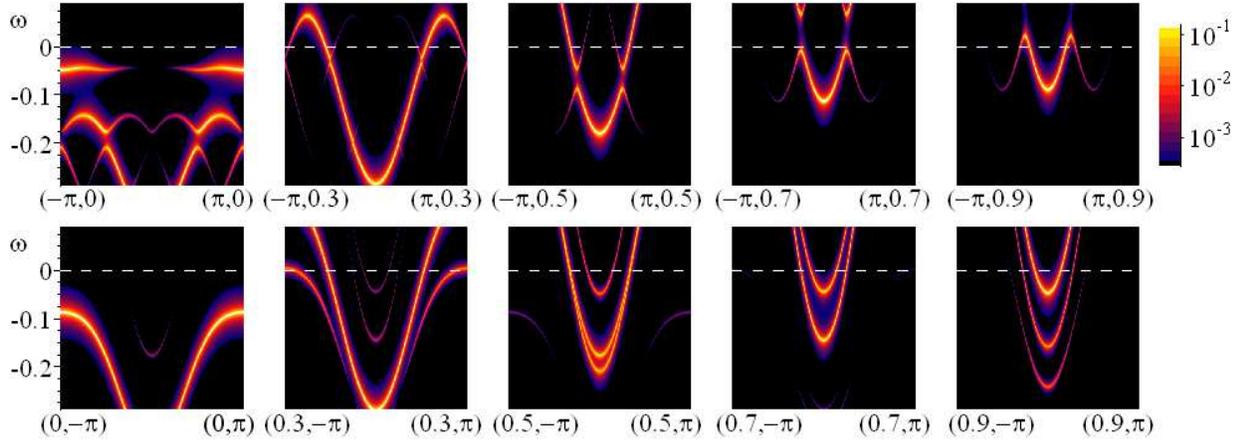}
\caption{
Momentum-space cuts of $A(k,\omega)$ along
horizontal (upper panel) and vertical (lower panel) lines in the Brillouin zone
for ideal vertical stripes,
with parameters as in Fig.~\ref{fig:pes1}a, left panel.
A spectral weight transfer between different subbands is clearly visible
in the lower panel, i.e. for cuts parallel to the stripe direction.
Note the hopping matrix element, $t=0.15$ eV, controls the energy scale on
the vertical axis; larger values may be appropriate to account for the
band structure on the scale of 1 eV.
}
\label{fig:cut1}
\end{figure*}

\section{Ordered stripes: Electronic spectra}
\label{sec:ord}

To set the stage, we start with results for perfectly ordered stripes.
We choose band structure parameters $t=0.15$ eV, $t'=-t/4$, $t''=t/12$.
Stripes are characterized by $\psi_x=(1+i)/\sqrt{2}$, $\kappa_1=0$,
$\kappa_2=-\kappa_3=0.03$ eV, giving a hopping modulation $\delta t$ of $\pm0.021$ eV.
Such ideally ordered stripes may be expected at temperatures much below the charge
ordering temperature in compounds with little quenched disorder and/or strong
commensurate lattice pinning.
Candidates are \lbco\ and Eu-doped \lsco\ near 1/8 hole doping.

Fig.~\ref{fig:pes1} shows Fermi surface (FS) cuts of the spectral function
$A({\vec k},\omega\!=\!0)$ in the normal state ($\Delta_0\!=\!0$).
The left panel has $\mu=-0.12$ eV, resulting in a hole doping of roughly 0.12,
whereas the right panel has $\mu=-0.02$ eV, illustrating the
effect of ``nested'' antinodal FS pieces (i.e. separated by $K_x$).
Parts a)--d) represent the same data, but differently:
a) displays the signal for vertical stripes with high energy and momentum
resolution and a logarithmic intensity color scale,
b) has reduced resolution,
c) shows the signal of horizontal and vertical stripes superimposed, and
d) shows the same on a linear intensity scale.
This comparison illustrates that the presence of both horizontal and vertical
stripes makes the identification of stripe signatures difficult,
and that shadow band features are present, but generically weak
due to matrix element effects.

\begin{figure}[b]
\centering
\includegraphics[width=3.1in]{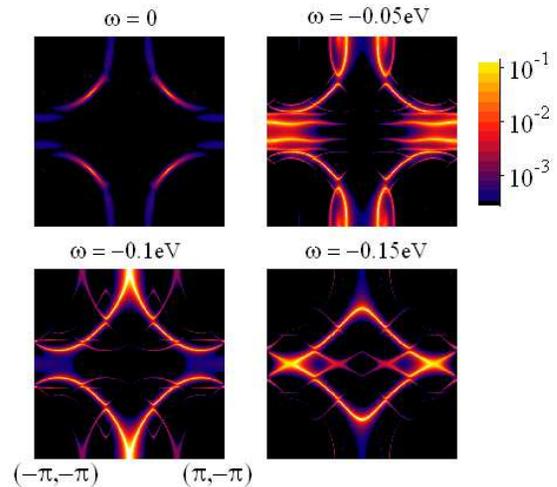}
\caption{
Constant-energy cuts at different energies through $A({\vec k},\omega)$
of ideal superconducting stripes with parameters
as in Fig.~\ref{fig:pes1} (left), but $\Delta_0=24$ meV,
$-\kappa_4=\kappa_5=7.5$ meV.
}
\label{fig:pes3}
\end{figure}

In Fig.~\ref{fig:cut1}, we display $A({\vec k},\omega)$ as function of
energy along horizontal and vertical cuts in momentum space,
for vertical stripes with parameters as in Fig.~\ref{fig:pes1}a left.
Finally, Fig.~\ref{fig:pes3} shows constant-energy cuts through
$A({\vec k},\omega)$ at different energies below the Fermi level,
but now for superconducting stripes with
$\Delta_0=24$ meV, $-\kappa_4=\kappa_5=7.5$ meV.

Figs.~\ref{fig:pes1}, \ref{fig:cut1}, and \ref{fig:pes3} nicely illustrate all ARPES features arising
from stripes. Those are:
(i)   broken rotation symmetry in momentum space,
(ii)  multiple bands and gaps, arising from Bragg reflection at the boundaries of the reduced Brillouin zone,
with the gaps being smallest along the momentum-space diagonal due to the $d$-wave-like form factor,
(iii)  shadow bands shifted by the ordering wavevector, which are weak in most parts
of momentum space,
(iv)  straight FS segments appearing near the antinodes in the direction perpendicular
to the stripes,
(v)   nodal QP survive at the FS.
Thus, the presence of nodal QP cannot be taken as evidence against stripes.
Closed FS pockets, which exist for certain parameter combinations, are
essentially invisible due to the small intensity.
Antinodal FS pieces in the direction parallel to the stripes may
disappear in situations close to nesting, due to a Bragg gap opening.
Then, adding horizontal and vertical stripe signals leads to
arc-like FS pieces (Fig.~\ref{fig:pes1} right).
Otherwise, the antinodal FS pieces display rather little stripe-induced shift
in momentum space: The Fermi-momentum locations along $(0,\pi)$--$(\pi,\pi)$ and
$(\pi,0)$--$(\pi,\pi)$,
determined from the most intense bands in Fig.~\ref{fig:pes1}a left,
differ by only $\Delta k_F\approx\pi/20$.
Signatures of broken rotation symmetry are more pronounced e.g. at 0.15 eV below
the Fermi level, with an overall oval-shaped constant-energy contour.

Most features discussed here are qualitatively similar to earlier work
\cite{salkola,seibold,granath,orgad}.
The visible differences between bond- and site-centered
stripes are minimal, being restricted to some matrix elements.
Accordingly, our FS structure is related to that found
in recent work \cite{millis} geared towards understanding of the
quantum oscillation experiments,
with the difference that our treatment does not include static magnetism
(which is important in forming small FS pockets, but does no drastically
influence the overall ARPES signal as discussed above).

In our subjective judgement, the most prominent stripe features,
possibly detectable experimentally,
are the straight FS segments near the antinodes (Figs.~\ref{fig:pes1})
and the spectral weight transfer between the subbands as function
of $k_x$ (Fig.~\ref{fig:cut1}).
The latter feature implies that, upon changing $k_x$, different
high-intensity subbands cross the Fermi level -- these
bands constitute the near-nodal FS pieces and the straight antinodal
pieces, respectively.


\section{Fluctuating stripes}

Modelling disordered stripes requires to account for fluctuations of $\psi_{x,y}$.
Those we assume to be described by a $\psi^4$-type theory $\mathcal{S}_\psi$
for the O(4) field $\psi = (\psi_x,\psi_y)$ \cite{maestro,vvk,robertson}.
The precise form of $\mathcal{S}_\psi$ will determine the character of the
fluctuations (amplitude vs. phase).
As argued elsewhere \cite{mv08,vvk},
a reasonable assumption is that amplitude fluctuations of $\psi$ are small,
i.e., stripe disordering is dominated by dislocations and domain walls.
Hence, we resort to a form of $\mathcal{S}_\psi$ with suppressed amplitude fluctuations:
\begin{eqnarray}
&&\mathcal{S}_{\psi} = \int d \tau d^2 {\vec r} \Bigl[
\left| \partial_\tau \psi_x \right|^2 +
\left| \partial_\tau \psi_y \right|^2 +
s_x |\psi_x|^2 + s_y |\psi_y|^2 \label{psiact}
\\
&&+
 c_{1x}^2 \left| \partial_x \psi_x \right|^2 + c_{2x}^2 \left|
\partial_y \psi_x \right|^2 + c_{1y}^2 \left|
\partial_y \psi_y \right|^2 + c_{2y}^2 \left|
\partial_x \psi_y \right|^2
\nonumber \\
&&+
u_1 \psi^4 + u_2 \psi^6
+ v |\psi_x|^2 |\psi_y|^2
+ w \left( \psi_x^4 \!+\! \psi_x^{\ast 4}
\!+\! \psi_y^4 \!+\! \psi_y^{\ast 4} \right) \Bigr]
\nonumber
\end{eqnarray}
with $\psi^2\!\equiv\!|\psi_x|^2\!+\!|\psi_y|^2$.
A combination of $u_1\!<\!0$ and $u_2\!>\!0$ suppresses amplitude fluctuations
of $\psi$.
The quartic $v |\psi_x|^2 |\psi_y|^2$ term regulates the repulsion or
attraction between horizontal and vertical stripes;
we shall employ $v\!>\!0$ leading to stripe-like order
(whereas $v\!<\!0$ results in checkerboard structures).
The phase-sensitive $w$ term provides commensurate pinning and
selects bond-centered (instead of site-centered)
stripes for $w\!>\!0$ \cite{kohsaka,vvk}.
(Small incommensurabilities enhance phase fluctuations,
but do not qualitatively change our results.)


\section{Disordered stripes: Electronic spectra}
\label{sec:dis}

A full treatment of the action $\mathcal{S}_c\!+\!\mathcal{S}_{\psi}\!+\!\mathcal{S}_{c\psi}$
is computationally difficult.
As in Refs.~\cite{mv08,vvk}, we adopt the adiabatic approximation
that the collective CDW field $\psi$ does not fluctuate on the time scale of the $c$
dynamics, being justified for disorder-pinned stripes.
Practically, lattice Monte Carlo (MC) simulations of $\mathcal{S}_{\psi}$ on lattices
with $64^2$ sites
are used to generate configurations of $\psi_{x,y}$, for each configuration
$\mathcal{S}_c\!+\!\mathcal{S}_{c\psi}$ is diagonalized to obtain the
electronic spectrum, which is then averaged over 50 configurations.
The underlying picture of short-range-ordered static stripes accounts
for the existence of stripe segments, checkerboard domain walls etc.,
but neglects stripe dynamics and inelastic processes.

\begin{figure}
\centering
\includegraphics[width=3.1in]{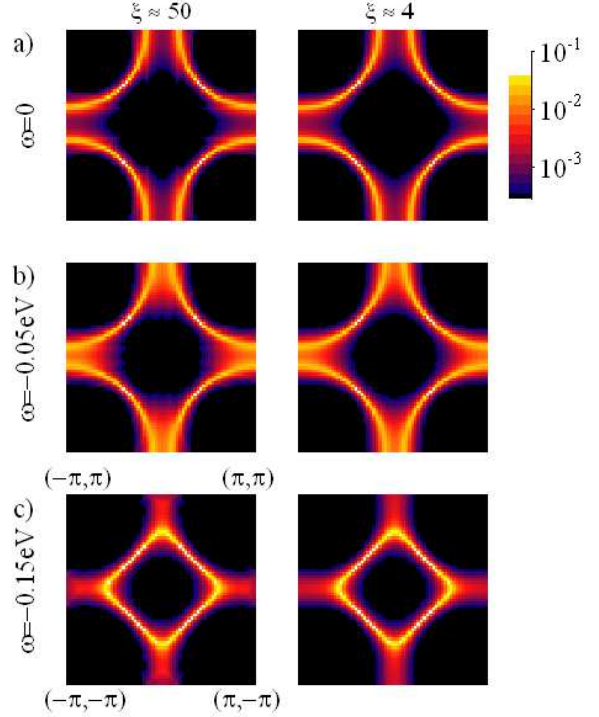}
\caption{
Constant-energy cuts through $A({\vec k},\omega)$
of short-range ordered non-superconducting stripes,
with band parameters as in Fig.~\ref{fig:pes1}.
The differences between stripe correlation lengths of
$\xi\approx 50$ (left) and $\xi\approx 4$ (right) are minimal,
and stripe features are barely visible.
}
\label{fig:pes4}
\end{figure}

\begin{figure}
\centering
\includegraphics[width=3.1in]{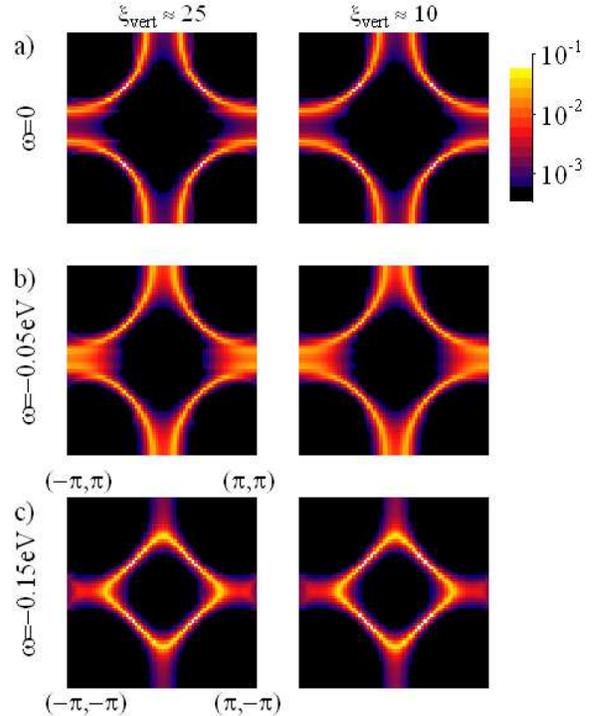}
\caption{
Same as in Fig.~\ref{fig:pes4}, but for short-range ordered stripes
in the presence of a uniaxial anisotropy.
}
\label{fig:pes5}
\end{figure}

To avoid proliferation of parameters, we do not explicitly model the impurity
pinning. Instead we simply work in a regime of $\mathcal{S}_{\psi}$ with short-range
order, and assume that the configuration average is equivalent to a disorder average.
[We expect that details of the impurity potential only matter for short-range
(i.e. point-like) impurities \cite{kaulimp}, whereas a smooth impurity potential
arising from out-of-plane defects essentially pins existing domains.]

Figs.~\ref{fig:pes4} and \ref{fig:pes5} show ARPES spectra for different $\xi$ \cite{param},
for situations without and with in-plane anisotropy [induced by slightly anisotropic
masses, $s_{x,y}$, in Eq.~(\ref{psiact})].
Note that $\xi\to\infty$ corresponds to the stripe ordering transition
where fluctuations are not small.

Without anisotropy, Fig.~\ref{fig:pes4}, stripe signatures are only discernible for
very large stripe domains. In contrast, for correlation lengths of 30 and below,
those signatures are smeared, and the strong scattering due to the stripe
disorder potential leads to rather incoherent antinodal QP.

The situation is different in the presence of an in-plane anisotropy, Fig.~\ref{fig:pes5}:
Here, stripe signatures (in particular the straight FS pieces near the antinodes)
are visible, provided that the stripes in the dominant direction have a correlation
length of order 10 or larger.
(Both cases in Fig.~\ref{fig:pes5} have an amplitude ratio
$\langle\psi_x^2\rangle/\langle\psi_y^2\rangle$ of roughly 2:1, i.e., vertical
stripes clearly dominate.)


\section{Stripes vs. nematics}

In detwinned samples of underdoped \ybco, signatures of spontaneous breaking of
rotation symmetry have been found in neutron scattering \cite{nematic}.
The spin fluctuation spectrum has been modelled assuming a $d$-wave nematic
instability \cite{yamase}.
However, it is also conceivable that the primary cause of rotation symmetry breaking
is the tendency toward stripe order (which may be preceded by a nematic transition).
The difference in these concepts is in the driving mechanism of the phenomenon.

We propose that the two situations can be distinguished in ARPES.
Fig.~\ref{fig:pes1} shows that the horizontal and vertical antinodal Fermi wavevectors
for ``realistic'' unidirectional stripes differ only by roughly $\pi/20$.
In a purely nematic picture, the order parameter is the effective anisotropy in the
hopping matrix elements $t_x$ and $t_y$, distorting the FS.
The difference between the horizontal and vertical $k_F$ is linear
in $\Delta t = (t_x-t_y)/2$, and $\Delta t/t \approx 4\%$ is sufficient
to produce an antinodal $\Delta k_F$ of $\pi/20$.
However, in Ref.~\cite{yamase} the effective $\Delta t/t$ required to model the neutron
data was about 20\%, which does not only shift the antinodal $k_F$, but
even leads to a change in the FS topology.

Independent of precise numbers, it is clear that ARPES will be able
to distinguish whether the driving mechanism of the anisotropy seen
in Ref.~\cite{nematic} is of stripe nature (leading to little antinodal shifts)
or of purely nematic nature (implying large antinodal shifts).


\section{Conclusion}

We have discussed photoemission signatures of
ordered and statically disordered valence-bond stripes in cuprates,
using a combination of mean-field theory for the
fermionic sector and Monte-Carlo simulations for
the stripe order parameter.
The results show clear stripe fingerprints for
perfectly ordered stripes, in qualitative agreement with earlier work.
These fingerprints should be visible in high-resolution experiments
on stripe-ordered 214 compounds.
However, moderate stripe disorder, in combination with the simultaneous presence
of horizontal and vertical stripe domains, is able to wash out most stripe
signatures.
This implies that stripes with a spatial correlation length
of e.g. 10 lattice spacings, which are easy to see in
an STM experiment, leave little trace in ARPES.
We have proposed that the position of antinodal quasiparticles
can be experimentally used to distinguish whether
rotation symmetry breaking in \ybco\ is driven by
stripe or nematic order.



This research was supported by the DFG through
SFB 608 (K\"oln) and Research Unit FG 538.
We thank S. Borisenko, B. B\"uchner,
J. C. Davis, J. Fink, M. Granath, K. Yamada,
and A. Yazdani
for discussions.


\end{document}